\documentclass[a4paper]{article}

\usepackage{INTERSPEECH2020}
\usepackage{multirow}
\usepackage{url}
\usepackage{float}
\usepackage{hyperref}
\hypersetup{
    colorlinks=true,
    urlcolor=black,
    citecolor=black,
    linkcolor=black
}

\title{Deep Speech Inpainting of Time-frequency Masks}
\name{Mikolaj Kegler$^{\star,1,2}$ \qquad Pierre Beckmann$^{\star,1,3}$ \qquad Milos Cernak$^{1}$\thanks{${\star}$-These authors contributed equally to this work.}}

\address{$^{1}$Logitech Europe S.A., 1015, Lausanne, Switzerland\\
$^{2}$Imperial College London, SW7 2AZ, London, UK\\
$^{3}$Swiss Federal Institute of Technology Lausanne, 1015, Lausanne, Switzerland}
\email{mak616@ic.ac.uk, pierre.beckmann@epfl.ch, milos.cernak@ieee.org}

\begin{document}

\maketitle
\begin{abstract}
Transient loud intrusions, often occurring in noisy environments, can completely overpower speech signal and lead to an inevitable loss of information. While existing algorithms for noise suppression can yield impressive results, their efficacy remains limited for very low signal-to-noise ratios or when parts of the signal are missing. To address these limitations, here we propose an end-to-end framework for speech inpainting, the context-based retrieval of missing or severely distorted parts of time-frequency representation of speech. The framework is based on a convolutional U-Net trained via deep feature losses, obtained using speechVGG, a deep speech feature extractor pre-trained on an auxiliary word classification task. Our evaluation results demonstrate that the proposed framework can recover large portions of missing or distorted time-frequency representation of speech, up to 400 ms and 3.2 kHz in bandwidth. In particular, our approach provided a substantial increase in STOI \& PESQ objective metrics of the initially corrupted speech samples. Notably, using deep feature losses to train the framework led to the best results, as compared to conventional approaches.
\end{abstract}

\section{Introduction}
\label{sec:introduction}

Recent major achievements in the field of speech enhancement (SE) have been mainly attributed to deep learning~\cite{xu2013experimental,wang2014training,weninger2015speech}. In particular, these new approaches tend to outperform traditional statistical SE systems, especially for high-variance noises~\cite{Kumar2016SpeechNetworks,Wang2017SupervisedOverview}. SE algorithms based on deep neural networks (DNNs) typically belong to one of the following groups: (i) causal systems that maintain the conventional approach of speech and noise estimation for spectral subtraction methods~\cite{xu2015regression, Mirsamadi2016,valin2018hybrid} and (ii) end-to-end systems, including generative approaches, that are usually non-causal and require longer temporal integration windows ~\cite{Pascual2017,Pascual2019TowardsNetworks,Liu2019}. Both approaches are capable of suppressing non-stationary noise, even at relatively low signal-to-noise ratios (SNRs). However, they rarely consider cases involving transient high-amplitude noise intrusions, which may completely overpower the speech signal. In such cases, the corrupted portion of the signal is inevitably lost and can only be restored.

To address the existing knowledge gap, here, we consider the task of speech inpainting, the context-based recovery of missing or severely degraded information in time-frequency representation of natural speech. The problem of generative, context-based, information recovery has been traditionally investigated in the field of computer vision, whereby it's referred to as image completion or inpainting~\cite{Iizuka2017GloballyCompletion, Yu2018GenerativeAttention, Liu2018ImageConvolutions}. A similar problem was formulated in the field of audio processing and named, by analogy, audio inpainting.
The problem was originally studied with only short time gaps~\cite{adler2011audio}. More recent approaches reported new promising results but only with music and musical instruments~\cite{Perraudin2016InpaintingGraphs, Marafioti2018AInpainting}, not natural speech.
Building on the existing audio inpainting, here, we introduce an end-to-end \textit{speech inpainting} system for recovering missing or severely degraded parts of time-frequency representation of speech.

We simulated the degradation of speech signals by masking their time-frequency representations. The applied masks represented either (i) time gaps, similar to packet loss~\cite{perkins1998survey,Lee2016PacketTransmission}, but with long intrusions of up to 400 ms in duration, (ii) frequency gaps, related to the bandwidth extension problem \cite{Iser2008BandwidthSignals, nagel2009harmonic} but with missing frequency bins summing up to 3200 Hz in bandwidth or (iii) irregular, random gaps disrupting up to 40\% of the overall time-frequency representation of 1-second-long speech segments. According to the authors' knowledge, this problem, at such scale, was not investigated in the past. Notably, a SE system that could recover missing or severely degraded parts of time-frequency representations of speech, of arbitrary shapes, has not been proposed to date.

To tackle the problem of speech inpainting we used an end-to-end DNN with the U-Net architecture~\cite{Ronneberger2015U-Net:Segmentation}. According to recent studies, the use of deep feature losses for training SE systems can improve their overall performance, depending on the feature extraction approach~\cite{Liao2019,Germain2019,Sahai2019SpectrogramSeparation}. We hypothesized that a specialized feature extractor, tailored specifically for speech processing, would provide the best performance of the trained system. Thus, we employed speechVGG~\cite{beckmann2019speech}, a deep speech feature extractor based on the classic VGG-16 architecture~\cite{Simonyan2015VERYRECOGNITION} and pre-trained on an auxiliary word classification task. We considered two configurations of the framework for \textit{informed} and \textit{blind} inpainting, depending on the availability of masks indicating missing or degraded parts of time-frequency representation of speech. In the case of blind speech inpainting, we evaluated the system using different types of intrusions, by replacing or adding high-amplitude noise to the masked parts of time-frequency representation of speech. Performance of the proposed approach was compared to a baseline based on the linear predictive coding (LPC) extrapolation algorithm~\cite{kauppinen2002audio}.

\section{Materials \& methods}
\label{sec:methods}

\subsection{Data \& preprocessing}
\label{sec:experiment}

We performed all of our experiments using LibriSpeech corpus, the open dataset of read English speech sampled at 16 kHz~\cite{Panayotov2015Librispeech:Books}. We used the \textit{train-clean-100} subset to train, \textit{test-clean} subset to monitor the training performance and \textit{dev-clean}, as a held-off data to evaluate all the models explored in this work. All available speech recordings were chunked into 1024-ms-long segments, without overlap. The time-frequency representation of each segment was obtained using a complex short-time Fourier transform (STFT, 256 samples window with 128 samples overlap, 128 frequency bins), resulting in a 128 x 128 matrix. The log-magnitude of each time-frequency bin was obtained by taking an absolute value of a complex number and computing its natural logarithm, and the phase, by computing its angle. Log-magnitude of each frequency channel was normalized using the mean and standard deviation obtained from the training data.

\subsection{Speech inpainting task}
\label{sec:TFmasks}

\begin{figure}[htbp]
\begin{minipage}[b]{\linewidth}
\centering
  \includegraphics[width=\linewidth]{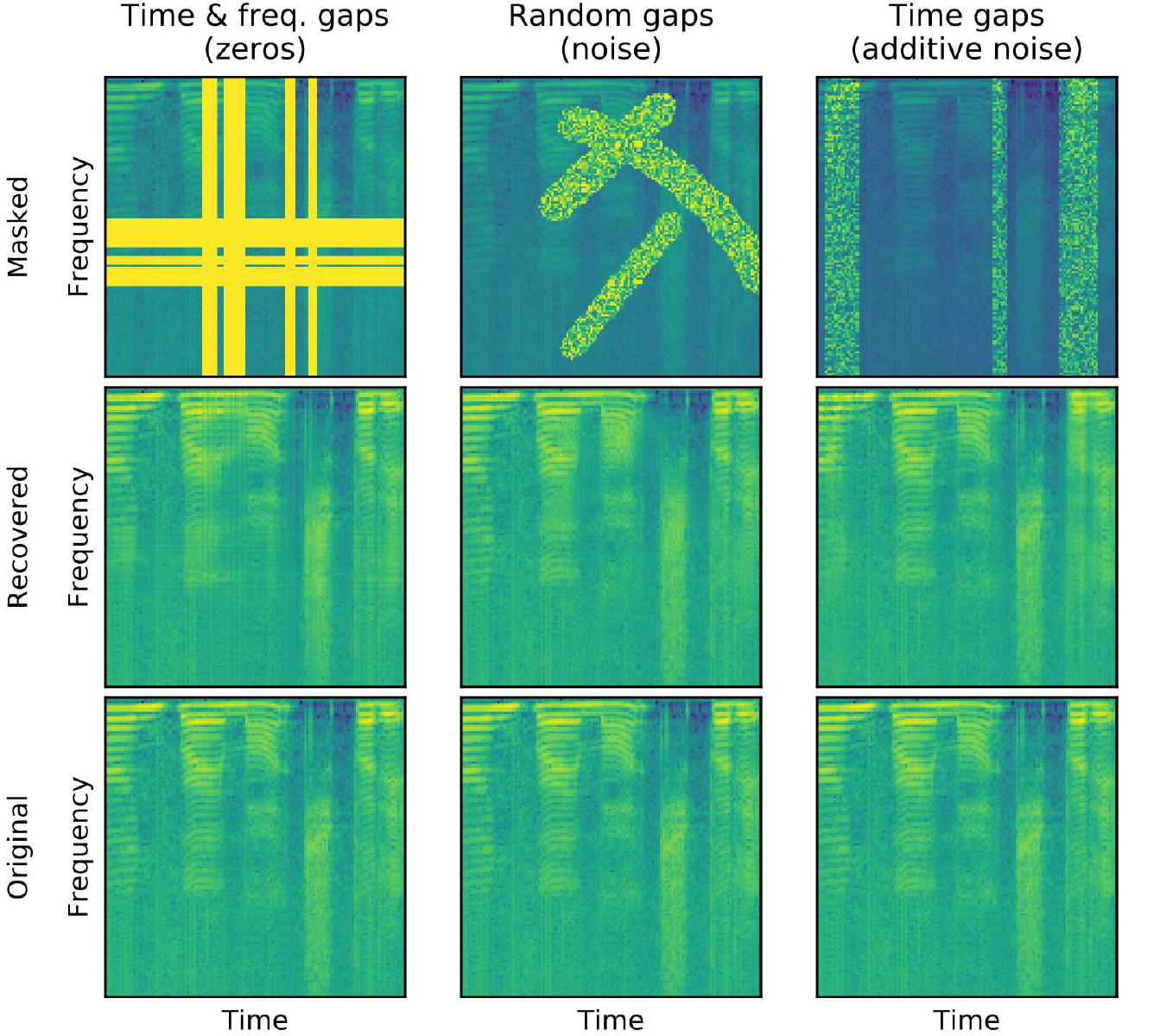}
    \caption{Examples of speech inpainting. Masked parts of STFT log-magnitudes of speech are either removed (left) or replaced (middle), as well as, mixed with high-amplitude noise (right).}
  \label{fig:reconstructions}
\end{minipage}
\end{figure}

In the task of speech inpainting, we simulated the degradation of speech signal by applying masks to the STFTs, obtained as specified in section~\ref{sec:experiment}. We considered three types of masks covering a specific portion of time-, time \& frequency information, or a random arbitrary combination of the two~(Fig.~\ref{fig:reconstructions}). For time and time \& frequency masks, the percentage indicated the masked portions of time and frequency bins, and in the random case, the overall mask coverage (i.e. area of the STFT). Each mask was randomly distributed between 1 to 4 blocks covering time and frequency bins, none shorter than 3 bins (24 ms or 187.5 Hz bandwidth). Number of blocks and their placement were each time drawn from uniform distributions. The mask was applied equivalently to the STFT magnitude and phase.

We trained our speech inpainting system to recover speech samples (from \textit{train-clean-100}) distorted using time \& frequency or random masks. For each training sample, the type of mask was assigned randomly with equal probability. Size of each mask used in training was drawn from the normal distribution~$N(\mu=29.4\%, \sigma=9.9\%)$. The speech inpainting performance of the trained system was evaluated using the held-off \textit{dev-clean} data and mask sizes ranging from 10\% (total: 100 ms, 800 Hz bandwidth), up to 40\% (total: 400 ms, 3200 Hz bandwidth). Speech samples from the held-off evaluation data were masked, processed through the trained speech inpainting system and reconstructed back to time-domain (i.e. waveform). The short term objective intelligibility (STOI) \cite{Taal2010ASpeech} and perceptual evaluation of speech quality (PESQ) \cite{PESQ1998}, measured between the processed and the original (i.e. non-masked) speech samples, were used to quantify the inpainting performance.

\begin{figure}[htbp]
\centering
  \includegraphics[width=\linewidth]{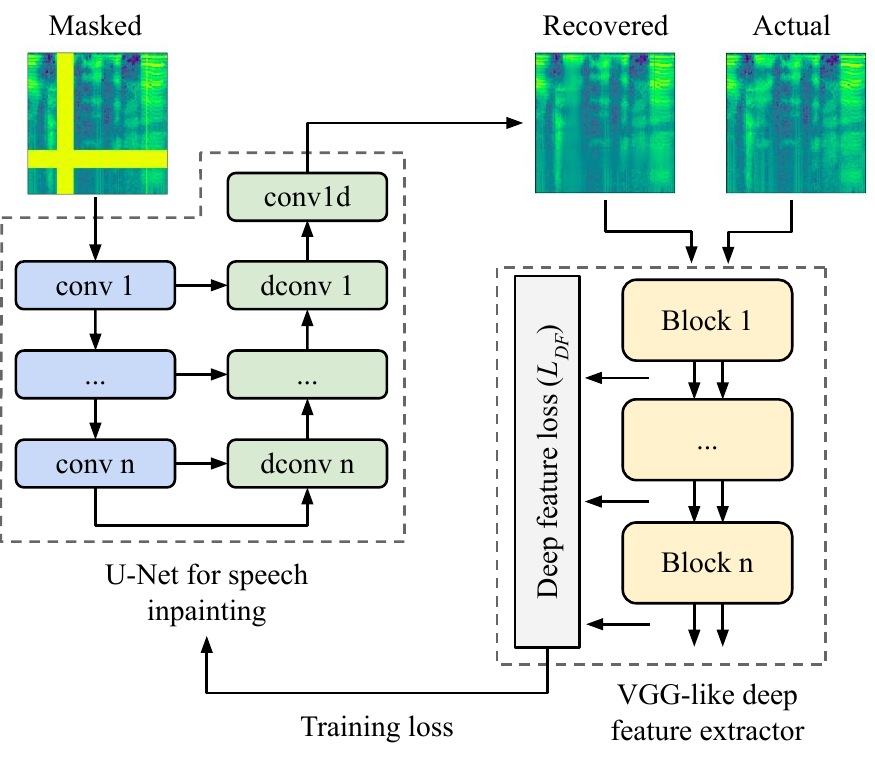}
  \caption{The speech inpainting framework is composed of the U-Net for speech inpainting (left, see~\ref{sec:UNet} for details) and VGG-like deep feature extractor (right). Deep feature losses $L_{DF}$ for training the U-Net are obtained by computing the $L_1$ distance between the representations of the recovered and the actual STFT log-magnitudes at pooling layers concluding each block of the feature extractor (see \ref{sec:DFL} for details).\vspace{-8pt}}
\label{fig:architecture}
\end{figure}

\subsection{U-Net for speech inpainting}
\label{sec:UNet}

The main part of the proposed speech inpainting framework was an end-to-end DNN with a U-Net architecture~\cite{Ronneberger2015U-Net:Segmentation}. The DNN was applied to restore missing or degraded parts of the log-magnitude of speech STFTs (Fig.~\ref{fig:reconstructions}), obtained, as describe in sections~\ref{sec:experiment}~\&~\ref{sec:TFmasks}. The phase information was discarded and not used by our model. To reconstruct the processed sample back to time domain, its phase was estimated directly from the magnitudes via the local weighted sums algorithm~\cite{LeRoux2010DAFx09,LeRoux2010ASJ09}.

The diagram of the system is presented in Fig.~\ref{fig:architecture}. Our U-Net was composed of six encoding (blue, Fig.~\ref{fig:architecture}) and seven decoding blocks (green, Fig.~\ref{fig:architecture}). Each encoding block consisted of a 2D convolution layer with a stride 2 and ReLU activation. In turn, each decoding block upsampled its input through a 2D upsampling layer with a kernel size 2, effectively doubling its size. Such upsampled input was subsequently concatenated with the input to the corresponding encoding block. The concatenated set of features was then processed through a 2D convolution (stride 1) with leaky ReLU activation ($\alpha=0.2$)~\cite{xu2015empirical}. Batch normalization~\cite{Ioffe2015BatchShift} was applied after each 2D convolution layer across the network. The final decoding block was a 1D convolution with a linear activation function returning the restored version of the distorted input speech sample. Parameters of our U-Net's convolution layers are listed in Table \ref{tab:conv-params}.

We considered two versions of the framework: performing informed and blind speech inpainting. In the informed case, the mask corrupting the input was known and all 2D convolutions in the network were replaced with partial convolutions (\textit{PC})~\cite{Liu2018ImageConvolutions}, which processed only valid (non-masked) parts of their input and ignored the rest. We also wanted to explore whether the framework can simultaneously identify and restore missing or degraded parts of the input. In such case, all convolution layers in the U-Net performed standard, full, 2D convolutions (\textit{FC}) and we refer to such setup as the blind speech inpainting. Other than that, the two configurations of the network were identical.

\begin{table}[htbp]
\centering
\caption{Parameters of the model's convolution layers. All 2D convolutions across the model employed zero-padding (i.e. 'same' mode). Block 0 represents the final 1D convolution.}
\label{tab:conv-params}
\begin{tabular}{c|c|c}
 & \multicolumn{2}{c}{(kernel size, number of filters)} \\ \hline
\multicolumn{1}{c|}{Block} & Encoding - \textit{conv} & Decoding - \textit{dconv} \\ \hline
\multicolumn{1}{c|}{0} & - & (1,1) \\ \hline
\multicolumn{1}{c|}{1} & (7, 16) & (3, 1) \\ \hline
\multicolumn{1}{c|}{2} & (5, 32) & (3, 16) \\ \hline
\multicolumn{1}{c|}{3} & (5, 64) & (3, 32) \\ \hline
\multicolumn{1}{c|}{4} & (3, 128) & (3, 64) \\ \hline
\multicolumn{1}{c|}{5} & (3, 128) & (3, 128) \\ \hline
\multicolumn{1}{c|}{6} & (3, 128) & (3, 128)
\end{tabular}
\end{table}

\subsection{Deep feature loss training}
\label{sec:DFL}

We used speechVGG~\cite{beckmann2019speech} as a deep feature extractor for training the speech inpainting framework. The extractor was based on the VGG-16 convolutional DNN architecture~\cite{Simonyan2015VERYRECOGNITION}  and consisted of five main blocks (Fig.~\ref{fig:architecture}, yellow), each concluded by a max-pooling layer. We pre-trained the speechVGG to classify 1000 most frequent, at least four-letters-long, words from the training data. The samples for pre-training were processed as specified in section \ref{sec:experiment}, augmented using SpecAugment \cite{Park2019SpecAugment:Recognition} and randomly padded with zeros to a size of 128 x 128.

We trained the framework for speech inpainting using deep feature losses $L_{DF}$~\cite{Liao2019,Germain2019,Sahai2019SpectrogramSeparation} (Fig.~\ref{fig:architecture}, grey), computed via the pre-trained speechVGG feature extractor. For each training sample, the U-Net was applied to recover the degraded input. The reconstructed ($\hat{Y}$) and the original ($Y$) log-magnitude STFTs were then processed through the speechVGG extractor. Activation of all five of the extractor's pooling layers $E$, one at the end of each block (Fig.~\ref{fig:architecture}, yellow) were obtained for the reconstructed $E(\hat{Y})$ and actual $E(Y)$ samples. Subsequently, the deep feature loss $L_{DF}$ was computed as the $L_1$ loss between the two representations:
\begin{equation}
    L_{DF} = L_{1}(E(Y), E(\hat{Y}))
\end{equation}
We compared the performance of the inpainting framework trained with deep feature losses obtained through: (i) the pre-trained speechVGG extractor (\textit{speechVGG}), (ii) the original VGG-16 network pre-trained to classify images from the ImageNet dataset~\cite{Simonyan2015VERYRECOGNITION, deng2009imagenet} (\textit{imgVGG}) or (iii) without deep feature losses, but the direct $L_1$, per-pixel loss, between the original and the recovered STFT log-magnitudes: $L_{1}(Y, \hat{Y})$ (\textit{noVGG}).

\subsection{Linear predictive coding baseline}
\label{sec:baeline}

Inspired by Marafioti, et al.~\cite{Marafioti2018AInpainting}, we compared the performance of different configurations of the proposed deep speech inpainting framework (see section~\ref{sec:DFL}) to the well-established LPC extrapolation algorithm~\cite{kauppinen2002audio}. The LPC was applied in the time domain, to recover missing parts of speech samples, given the signal surrounding each intrusion. While the LPC method was suitable for known and well-defined, temporal intrusions, we couldn't find a meaningful way to apply it for recovering samples distorted with random irregular masks of arbitrary shapes. This highlights the flexibility of our end-to-end system, as compared to conventional time-domain approaches~\cite{Marafioti2018AInpainting,Lee2016PacketTransmission,kauppinen2002audio}.

\subsection{Implementation details}
The speechVGG was pre-trained using cross-entropy loss for 50 epochs with a fixed learning rate set to $5\times10^{-5}$. Each considered configuration of the U-Net for speech inpainting was trained for 30 epochs using either deep feature- ($L_{DF}$) or per-pixel- ($L_1$) loss with a fixed learning rate of $2\times10^{-4}$. ADAM optimizer~\cite{Kingma2014Adam:Optimization} was used in all the training routines.

\section{Results}
\label{sec:Results}

\subsection{Informed speech inpainting}
\label{sec:informed}

\begin{table*}[htbp]
\centering
\caption{Informed speech inpainting (section~\ref{sec:informed}). STOI \& PESQ were computed between the recovered and the actual speech segments from the validation set and averaged. The best scores were denoted in bold. See section \ref{sec:DFL} for training configurations. \vspace{-6pt}}
\label{tab:partial_conv}
\resizebox{0.88\textwidth}{!}{
\begin{tabular}{cccc|cc|cc|cc|cc|cc}
 &  & \multicolumn{2}{|c|}{Gaps} & \multicolumn{2}{c|}{Noise} & \multicolumn{2}{c|}{LPC} & \multicolumn{2}{c|}{noVGG} & \multicolumn{2}{c|}{imgVGG} & \multicolumn{2}{c}{speechVGG} \\
\multicolumn{1}{c|}{Intrusion} & \multicolumn{1}{c|}{Size} & STOI & PESQ & STOI & PESQ & STOI & PESQ & STOI & PESQ & STOI & PESQ & STOI & PESQ \\ \hline
\multicolumn{1}{c|}{\multirow{4}{*}{Time}} & \multicolumn{1}{c|}{10\%} & 0.893 & 2.561 & 0.901 & 2.802 & 0.921 & 2.798 & 0.926 & 3.118 & 0.917 & 3.117 & \textbf{0.938} & \textbf{3.240} \\
\multicolumn{1}{c|}{} & \multicolumn{1}{c|}{20\%} & 0.772 & 1.872 & 0.800 & 2.260 & 0.842 & 2.483 & 0.879 & 2.755 & 0.860 & 2.713 & \textbf{0.887} & \textbf{2.809} \\
\multicolumn{1}{c|}{} & \multicolumn{1}{c|}{30\%} & 0.641 & 1.476 & 0.688 & 1.919 & 0.750 & 2.233 & 0.805 & 2.428 & 0.779 & 2.382 & \textbf{0.811} & \textbf{2.450} \\
\multicolumn{1}{c|}{} & \multicolumn{1}{c|}{40\%} & 0.536 & 1.154 & 0.598 & 1.665 & 0.669 & 2.015 & 0.724 & 2.171 & 0.695 & 2.109 & \textbf{0.730} & \textbf{2.179} \\ \hline
\multicolumn{1}{c|}{\multirow{4}{*}{\begin{tabular}[c]{@{}c@{}}Time\\ +\\ Freq.\end{tabular}}} & \multicolumn{1}{c|}{10\%} & 0.869 & 2.423 & 0.873 & 2.575 & 0.887 & 2.692 & 0.905 & 2.921 & 0.899 & 2.915 & \textbf{0.920} & \textbf{3.034} \\
\multicolumn{1}{c|}{} & \multicolumn{1}{c|}{20\%} & 0.729 & 1.790 & 0.746 & 2.010 & 0.780 & 2.378 & 0.845 & 2.518 & 0.829 & 2.491 & \textbf{0.853} & \textbf{2.566} \\
\multicolumn{1}{c|}{} & \multicolumn{1}{c|}{30\%} & 0.598 & 1.391 & 0.629 & 1.653 & 0.668 & 2.128 & 0.765 & 2.158 & 0.743 & 2.134 & \textbf{0.772} & \textbf{2.178} \\
\multicolumn{1}{c|}{} & \multicolumn{1}{c|}{40\%} & 0.484 & 1.053 & 0.520 & 1.329 & 0.566 & \textbf{1.907} & 0.672 & 1.840 & 0.644 & 1.809 & \textbf{0.680} & 1.845 \\ \hline
\multicolumn{1}{c|}{\multirow{4}{*}{Random}} & \multicolumn{1}{c|}{10\%} & 0.880 & 2.842 & 0.892 & 3.063 & \multicolumn{2}{c|}{\multirow{4}{*}{N/A}} & 0.941 & 3.477 & 0.927 & 3.399 & \textbf{0.944} & \textbf{3.496} \\
\multicolumn{1}{c|}{} & \multicolumn{1}{c|}{20\%} & 0.809 & 2.233 & 0.830 & 2.543 & \multicolumn{2}{c|}{} & 0.912 & 3.079 & 0.897 & 3.040 & \textbf{0.918} & \textbf{3.114} \\
\multicolumn{1}{c|}{} & \multicolumn{1}{c|}{30\%} & 0.713 & 1.690 & 0.745 & 2.085 & \multicolumn{2}{c|}{} & 0.872 & 2.702 & 0.856 & 2.680 & \textbf{0.878} & \textbf{2.735} \\
\multicolumn{1}{c|}{} & \multicolumn{1}{c|}{40\%} & 0.644 & 1.355 & 0.682 & 1.802 & \multicolumn{2}{c|}{} & 0.837 & 2.443 & 0.823 & 2.422 & \textbf{0.846} & \textbf{2.479}
\end{tabular}%
}
\end{table*}

The first experiment assessed the performance of the proposed framework in the informed speech inpainting task when the exact masks were known. Here, masked parts of the inputs were replaced with zeros, reflecting missing information and the system was using partial convolutions~\cite{Liu2018ImageConvolutions}, which ignored masked parts of their inputs. Unprocessed, corrupted speech samples (\textit{Gaps}) and the case when their masked parts were filled with speech-shaped noise (\textit{Noise}), derived from the original speech signal, served as a reference to compare other methods against.

The complete evaluation results are presented in Table~\ref{tab:partial_conv}. Each configuration of the proposed framework improved both STOI and PESQ for all the considered shapes and sizes of intrusions, as compared to the distorted case with either zeros- or noise-filled gaps. The use of deep feature losses in training yielded better results, as compared to the case when the U-Net was trained via direct, per-pixel, $L_1$ loss (\textit{noVGG}). However, only the speechVGG (\textit{speechVGG}), not the VGG-16 pre-trained on the ImageNet data (\textit{imgVGG}), led to particularly notable improvements in STOI \& PESQ. The proposed framework trained with the speechVGG consistently outperformed the baseline LPC algorithm (\textit{LPC}) in terms of STOI, suggesting better recovery of speech intelligibility. For PESQ scores, the advantage of our method was consistent but smaller. Surprisingly, for the largest time-frequency masking (40\%) the LPC achieved substantially lower STOI, but it was the only case when it yielded slightly higher PESQ, as compared to our approach.

\subsection{Blind speech inpainting}
\label{sec:blind}

\begin{table*}[htbp]
\centering
\caption{Blind speech inpainting (section~\ref{sec:blind}). STOI \& PESQ were computed between the recovered and the actual speech segments from validation set and averaged. The best scores were denoted in bold. PC - informed (partial conv.), FC - blind (full conv.) inpainting.\vspace{-6pt}}
\label{tab:full-conv}
\resizebox{0.88\textwidth}{!}{
\begin{tabular}{cccc|cc|cc|cc|cc|cc}
 &  & \multicolumn{2}{|c|}{Gaps} & \multicolumn{2}{c|}{Noise} & \multicolumn{2}{c|}{PC - gaps} & \multicolumn{2}{c|}{FC - gaps} & \multicolumn{2}{c|}{FC - noise} & \multicolumn{2}{c}{FC - additive} \\
\multicolumn{1}{c|}{Intrusion} & \multicolumn{1}{c|}{Size} & STOI & PESQ & STOI & PESQ & STOI & PESQ & STOI & PESQ & STOI & PESQ & STOI & PESQ \\ \hline
\multicolumn{1}{c|}{\multirow{4}{*}{Time}} & \multicolumn{1}{c|}{10\%} & 0.893 & 2.561 & 0.901 & 2.802 & \textbf{0.938} & \textbf{3.240} & 0.930 & 3.191 & 0.919 & 3.066 & 0.933 & 3.216 \\
\multicolumn{1}{c|}{} & \multicolumn{1}{c|}{20\%} & 0.772 & 1.872 & 0.800 & 2.260 & 0.887 & 2.809 & 0.875 & 2.725 & 0.863 & 2.677 & \textbf{0.906} & \textbf{2.971} \\
\multicolumn{1}{c|}{} & \multicolumn{1}{c|}{30\%} & 0.641 & 1.476 & 0.688 & 1.919 & 0.811 & 2.450 & 0.798 & 2.384 & 0.792 & 2.374 & \textbf{0.882} & \textbf{2.788} \\
\multicolumn{1}{c|}{} & \multicolumn{1}{c|}{40\%} & 0.536 & 1.154 & 0.598 & 1.665 & 0.730 & 2.179 & 0.714 & 2.086 & 0.707 & 2.072 & \textbf{0.854} & \textbf{2.617} \\ \hline
\multicolumn{1}{c|}{\multirow{4}{*}{\begin{tabular}[c]{@{}c@{}}Time\\ +\\ Freq.\end{tabular}}} & \multicolumn{1}{c|}{10\%} & 0.869 & 2.423 & 0.873 & 2.575 & \textbf{0.920} & \textbf{3.034} & 0.911 & 3.000 & 0.896 & 2.859 & 0.912 & 3.023 \\
\multicolumn{1}{c|}{} & \multicolumn{1}{c|}{20\%} & 0.729 & 1.790 & 0.746 & 2.010 & 0.853 & 2.566 & 0.840 & 2.490 & 0.828 & 2.411 & \textbf{0.885} & \textbf{2.759} \\
\multicolumn{1}{c|}{} & \multicolumn{1}{c|}{30\%} & 0.598 & 1.391 & 0.629 & 1.653 & 0.772 & 2.178 & 0.757 & 2.108 & 0.743 & 2.041 & \textbf{0.854} & \textbf{2.562} \\
\multicolumn{1}{c|}{} & \multicolumn{1}{c|}{40\%} & 0.484 & 1.053 & 0.520 & 1.329 & 0.680 & 1.845 & 0.665 & 1.772 & 0.659 & 1.787 & \textbf{0.828} & \textbf{2.413} \\ \hline
\multicolumn{1}{c|}{\multirow{4}{*}{Random}} & \multicolumn{1}{c|}{10\%} & 0.880 & 2.842 & 0.892 & 3.063 & \textbf{0.944} & \textbf{3.496} & 0.932 & 3.442 & 0.917 & 3.272 & 0.932 & 3.435 \\
\multicolumn{1}{c|}{} & \multicolumn{1}{c|}{20\%} & 0.809 & 2.233 & 0.830 & 2.543 & \textbf{0.918} & 3.114 & 0.904 & 3.061 & 0.887 & 2.897 & 0.910 & \textbf{3.117} \\
\multicolumn{1}{c|}{} & \multicolumn{1}{c|}{30\%} & 0.713 & 1.690 & 0.745 & 2.085 & 0.878 & 2.735 & 0.869 & 2.701 & 0.853 & 2.596 & \textbf{0.891} & \textbf{2.893} \\
\multicolumn{1}{c|}{} & \multicolumn{1}{c|}{40\%} & 0.644 & 1.355 & 0.682 & 1.802 & 0.846 & 2.479 & 0.832 & 2.412 & 0.813 & 2.307 & \textbf{0.868} & \textbf{2.664}
\end{tabular}%
}
\end{table*}

In the blind speech inpainting, the masks disrupting the input speech were not known and the system was trained to simultaneously identify and recover distorted portions of the input. We considered three different scenarios by setting the masked values of the input STFT log-magnitudes to either zero (\textit{-gaps}, Fig.~\ref{fig:reconstructions}, left), white noise (\textit{-noise}, Fig.~\ref{fig:reconstructions}, middle) or a mixture of the original information and the noise (\textit{-additive}, Fig.~\ref{fig:reconstructions}, right). The noise was added directly to the STFT log-magnitudes and its amplitude was set to provide transient mixtures at the very low SNRs, below -10 dB (i.e. very disruptive noise level).

In this experiment, we used the best overall configuration of the system from the previous experiment, namely \textit{speechVGG}. The network setup remained the same, expect all partial convolutions (\textit{PC}) were replaced with regular, full 2D convolutions (\textit{FC}), as specified in section~\ref{sec:UNet}. The framework was re-trained and re-evaluated separately for each type of intrusion (\textit{-gaps}, \textit{-noise}, \textit{-additive}). Other than that, the training and evaluation routines remained the same as in the informed case.

The complete evaluation results are presented in Table~\ref{tab:full-conv}. All of the considered framework configurations successfully recovered missing or degraded parts of the input speech resulting in improved STOI and PESQ scores. The framework for the informed inpainting (\textit{PC-gaps}) yielded better results, as compared to its blind counterpart, when the masked parts of the input were set to either zeros or random noise (\textit{FC -gaps, -noise}, respectively). Notably, when input speech was mixed with, not replaced by, the high-amplitude noise, the framework operating in the blind setup (\textit{FC-additive}) led to the best results for larger intrusions of all the considered shapes. These results suggest that the framework for the blind speech inpainting takes advantage of the original information underlying the noisy intrusions, even when the transient SNR is very low (below -10 dB).

\section{Discussion}
\label{sec:discussion}

The proposed framework is capable of recovering missing or degraded parts of time-frequency representations of speech, as indicated by the substantial improvements in STOI and PESQ scores. Notably, intrusions used in the evaluation were as large as 400 ms (i.e. the duration of a syllable or a short word) and 3.2 kHz in bandwidth. We showed that employing deep feature losses in training leads to better results as compared to conventional methods. However, only the speech-specific feature extractor \textit{speechVGG} applied to obtain deep feature losses, unlike \textit{imgVGG} pre-trained on a visual task, led to the improved performance. Our results suggest that the proposed system can simultaneously identify degraded parts of the input speech and recover them. In particular, our system for blind speech inpainting improved STOI and PESQ scores of degraded speech, both when parts of the time-frequency representation were missing, as well as, distorted by the additive noise.

Importantly, our experiments employing additive noise represented only a simplified scenario. In future experiments, a broader range of ecological noises shall be considered to assess the framework performance in the joint speech denoising and inpainting task. Our approach could further benefit from incorporating phase information as an additional input feature for the DNN. We indeed experimented in this domain but none of our several attempts of utilizing the STFT phase in the framework outperformed the proposed approach. The limited phase reconstruction capacity may be underlying the only case when the LPC-based algorithm yielded higher PESQ score than our approach. Despite low STOI, the time-domain LPC reinforced phase consistency and although the recovered speech was unintelligible, its quality was quantified as disproportionately better.

\section{Conclusion}
\label{sec:concolusion}

We introduced a novel end-to-end framework for recovering large portions of missing or degraded time-frequency representation of natural speech. Our system provided substantial improvements in STOI \& PESQ metrics, both when the position of intrusions in time and frequency were known (informed-) or not (blind inpainting). The DNN at the core of our system benefited from being trained via deep feature losses. Computing the loss through speechVGG, a pre-trained speech feature extractor, led to the best results. We believe that the proposed approach can be integrated with the existing SE systems and contribute to the development of the next-generation general-purpose SE.

Demo of the deep speech inpainting is available at\footnote{\url{https://mkegler.github.io/SpeechInpainting}}. Python implementation of the speechVGG and pre-trained models are openly available at\footnote{\url{https://github.com/bepierre/SpeechVGG}}.

\section{Acknowledgements}
\label{sec:acknowledgements}

This work was funded by Logitech \& Imperial College London.

\bibliographystyle{IEEEtran}

\bibliography{references}

\end{document}